\documentstyle[12pt,epsf,graphicx]{article}
\def\href#1#2{#2}   
\newif\ifdraft
\draftfalse	  
\reversemarginpar   
\makeatletter
\let\mlabel=\label
\let\adkendequation=\endequation%
\def\endequation{\adkendequation\adklabel\global\@ignoretrue}
\let\adkendeqnarray=\endeqnarray%
\def\endeqnarray{\adkendeqnarray\adklabel\global\@ignoretrue}
\newbox\marglabbox
\def\adklabel{\ifvoid\marglabbox\else\marginpar{\unhbox\marglabbox}\fi}
\def\label#1{\ifdraft\ifmmode%
  \global\setbox\marglabbox=\hbox{\hfill\fbox{\tiny\verb*~#1~}}%
  \else\ifinner\else\marginpar{\hfill\fbox{\tiny\verb*~#1~}}%
  \fi\fi\fi \mlabel{#1}}
\makeatother
\ifdraft%
\fi
\font\twelvebb=msbm12
\font\tenbb=msbm10
\font\sevenbb=msbm7
  \newfam\bbfam
  
  \textfont\bbfam=\twelvebb
  \scriptfont\bbfam=\tenbb
  \scriptscriptfont\bbfam=\sevenbb
\font\twelveeusm=eusm10 scaled 1200
\font\teneusm=eusm10
  \newfam\eusmfam
  
  \textfont\eusmfam=\twelveeusm
  \scriptfont\eusmfam=\teneusm
  \scriptscriptfont\eusmfam=\scriptfont\eusmfam
\font\twelvefrak=eufm10 scaled 1200
\font\tenfrak=eufm10
  \newfam\frakfam
  
  \textfont\frakfam=\twelvefrak
  \scriptfont\frakfam=\tenfrak
  \scriptscriptfont\frakfam=\scriptfont\frakfam
\def\sqr#1#2{{\vcenter{\hrule height.#2pt
   \hbox{\vrule width.#2pt height#1pt \kern#1pt
      \vrule width.#2pt}
   \hrule height.#2pt}}}

\def\bsqr#1#2{{\vrule width #1pt height#2pt}}
\def\bsquare{{\mathchoice\bsqr66\bsqr66\bsqr33\bsqr33}}
\def\badbreak{\penalty1000}

\newcommand{\cE}{{\cal E}}                  

\begin{document}

\begin{center}
{\Large{\bf The Negativity of the Overlap-Based Topological}} \\
\vspace*{.1in}
{\Large{\bf Charge Density Correlator in Pure--Glue QCD}}\\
\vspace*{.1in}
{\Large{\bf and the Non-Integrable Nature of its Contact Part}}\\
\vspace*{.4in}
{\large{I.~Horv\'ath$^1$,
A.~Alexandru$^1$,
J.B.~Zhang$^2$,
Y.~Chen$^3$,
S.J.~Dong$^1$,
T.~Draper$^1$}}\\
\vspace*{.1in}
{\large{
K.F.~Liu$^1$,
N.~Mathur$^1$,
S.~Tamhankar$^1$ and
H.B.~Thacker$^4$}} \\
\vspace*{.15in}
$^1$Department of Physics and Astronomy, University of Kentucky, Lexington, KY 40506\\
$^2$CSSM and Department of Physics, University of Adelaide, Adelaide, SA 5005, Australia\\
$^3$Institute of High Energy Physics, Academia Sinica, Beijing 100039, P.R. China\\
$^4$Department of Physics, University of Virginia, Charlottesville, VA 22901

\vspace*{0.2in}
{\large{Apr 7 2005}}

\end{center}

\vspace*{0.15in}

\begin{abstract}
  \noindent
  We calculate the lattice two-point function of topological charge density in pure-glue QCD 
  using the discretization of the operator based on the overlap Dirac matrix. Utilizing data 
  at three lattice spacings it is shown that the continuum limit of the correlator complies
  with the requirement of non-positivity at non-zero distances. For our choice 
  of the overlap operator and the Iwasaki gauge action we find that the size 
  of the positive core is $\approx 2\,a$ (with $a$ being the lattice spacing) sufficiently 
  close to the continuum limit. This result confirms that the overlap-based topological charge 
  density is a valid local operator over realistic backgrounds contributing to the QCD path 
  integral, and is important for the consistency of recent results indicating the existence 
  of a low-dimensional global brane-like topological structure in the QCD vacuum. We also 
  confirm the divergent short-distance behavior of the correlator, and the non-integrable 
  nature of the associated contact part.

\end{abstract}

\noindent
{\bf 1. Introduction.} 
An intriguing property of the topological charge density (TChD) two-point function in 
Euclidean gauge theory has been pointed out long ago by Seiler and Stamatescu~\cite{SeSt}. 
Specifically, the correlator is non-positive at arbitrary non-zero distance. While not 
widely known or used, this fact arises straightforwardly as a consequence of 
reflection positivity and the pseudoscalar nature of the corresponding local field operator.  
There are (at least) two situations where this seemingly unusual property of the correlator
plays a relevant role. The first one involves the discussion of subtleties arising in the 
derivation of the Witten-Veneziano relation for the $\eta'$ mass~\cite{SeSt,WitVen}. Indeed, 
the topological susceptibility is positive by usual definition and yet, it can be equivalently 
expressed as a space-time integral of the correlator which is non-positive everywhere except 
at the origin. The expected non-integrable behavior of the (negative) correlator near the 
origin has to be countered by positive divergent terms (with support at the origin) to yield 
a finite positive susceptibility. This raises both legitimate conceptual issues about the role 
of short-distance fluctuations in the associated physics~\cite{SeSt}, as well as intriguing 
questions about how exactly does the cancellation of positive and negative infinities take place 
in the context of a lattice non-perturbative definition of the theory. Certain points related 
to these issues were discussed in the context of CP(N-1) models in Refs.~\cite{Vic99,Tha02}. 

The second instance where the negativity of TChD correlator has non-trivial implications
relates to questions about the nature of topological charge fluctuations in the QCD 
vacuum~\cite{Hor02B,Hor03A,Hor05A}. Indeed, if there exists a {\em fundamental structure}
in typical configurations contributing to the path integral of the theory with non-trivial 
ultraviolet behavior (such as QCD), then the space-time characteristics of such structure
should be consistent with the negativity of the TChD correlator.\footnote{Note that by 
fundamental structure we mean a structure that contains fluctuations 
at all scales and is in principle relevant for all aspects of QCD physics.} 
This requirement means, in particular, that such fundamental structure cannot be dominated 
by gauge fields supporting 4-dimensional sign-coherent regions of TChD~\cite{Hor02B,Hor03A}. 
On the other hand, the negativity of the correlator can be satisfied in an ordered manner 
if the structure involves interleaved layers of oppositely charged {\em lower-dimensional\/} 
regions. The existence of such low-dimensional brane-like structure in Monte Carlo generated 
lattice QCD configurations has been demonstrated~\cite{Hor03A}, 
\footnote{The low-dimensional nature of the fundamental topological field is reflected to 
some degree also in low-lying Dirac modes~\cite{Hor02A,Aub04} but the precise form of such 
correspondence is not known. The notion of strictly low-dimensional structure also emerged 
recently using indirect projection techniques~\cite{Zakh}. Its relation to the structure
in topological field is not clear at this point.} and it was shown that it 
behaves as an inherently global entity in the sense that its localized parts are not 
sufficient to explain the value of topological susceptibility in pure-glue 
QCD~\cite{Hor05A}.

A crucial ingredient for both of the above developments is the availability of a new kind 
of lattice TChD operator that can be used in the context of a non-perturbative definition 
of the theory. Indeed, the recent progress in putting the derivation of Witten-Veneziano 
formula on firmer ground~\cite{Giu02A} is based on the use of topological field associated 
with Ginsparg-Wilson fermions~\cite{Has98A,NarNeu95}. In fact, such a topological field 
exhibits properties analogous to those in the continuum and appears to lead to a satisfactory 
definition of topological susceptibility also in full QCD~\cite{GiuRosTes04,Lusch04}.
Similarly, the low-dimensional long-range topological structure has been observed using 
the operator based on the overlap Dirac matrix, and is not obviously visible when various 
naive operators are used~\cite{Hor03A}. The underlying reason leading to such niceties is 
tied to the fact that TChD operators based on chiral fermions appear to have a proper control 
of short-distance fluctuations. Indeed, the artificial ultraviolet infinities, present 
when naive lattice operators are used, apparently disappear with Ginsparg-Wilson TChD 
operators and no power-divergent subtractions to the susceptibility are needed. On the other 
side of the coin, the uncontrolled short-distance fluctuations of naive operators mask 
the presence of the ordered long-range topological structure in the QCD vacuum which however 
becomes apparent when these fluctuations are properly treated~\cite{Hor03Apr}.

Despite the special significance of TChD operators based on Ginsparg-Wilson fermions, the 
negativity of the correlator has not been numerically demonstrated for any particular choice
of the operator.\footnote{The early attempt to verify the negativity can be found in 
Ref.~\cite{Hor02Bpr}.} 
The reason why negativity is not obviously satisfied here has 
to do with the non-ultralocal nature of Ginsparg-Wilson fermions. Indeed, it was 
shown~\cite{nonultr} that if $D$ is any (otherwise acceptable) Ginsparg-Wilson operator, 
then $D_{x,y}$ is non-zero for arbitrarily large distances $|x-y|$. While not proved 
rigorously, it is expected that the analogous property holds also in terms of gauge variables 
in the sense that $D_{x,y}$ receives small but non-zero contributions from gauge paths that 
extend arbitrarily far away from $x$ and $y$. As a consequence, the TChD operator 
$q(x) \propto \mbox{\rm tr} \,\gamma_5 \, D_{x,x}$ is non-ultralocal in this sense
and cannot be strictly contained in any finite lattice region. This complication
makes direct arguments involving reflection positivity inexact at the lattice level 
even when the underlying lattice action is otherwise reflection positive. Nevertheless, 
assuming that $q(x)$ is a valid {\em local\/} lattice TChD operator, the consequences of 
reflection positivity are expected to hold upon taking the continuum limit. 
In this sense, verifying the negativity of the correlator at arbitrary non-zero physical 
distances represents a non-trivial check on the locality of the lattice operator
and on the consistency of the lattice action defining the theory (in cases where the
lattice action is not manifestly reflection positive, such as in full QCD with overlap
fermions). 

In what follows, we will verify numerically that the TChD operator constructed from overlap
Dirac matrix~\cite{Neu98BA} indeed leads to a negative correlator at non-zero physical 
distances in the continuum limit of pure-glue gauge theory. While our calculation is 
performed with a particular (but generic) choice of the overlap operator ($\rho=1.368$) 
and with a particular choice of ultralocal gauge action (Iwasaki action~\cite{Iwasaki}), 
we do not expect this conclusion to change for other generic choices. It is of some 
practical interest to quantify the size of the positive core $r_c^p$ of the correlator 
at the regularized level. We find that for lattice spacing $a=0.082$ fm the size is 
$r_c^p \approx 0.18$ fm. Using data at three different lattice spacings we obtain 
the continuum-extrapolated shape of the lattice correlator at small lattice distances. 
From this calculation we conclude that the size of the positive core sufficiently close 
to the continuum limit is $r_c^p\approx 2 a$ in our case (thus shrinking to zero
in a corresponding manner). Finally, we verify that 
the contribution of positive core of the correlator to susceptibility indeed diverges 
in the continuum limit as expected from general arguments. The nature of this divergence 
will be discussed quantitatively in an upcoming publication.

Before starting, we wish to emphasize a point of convention. Since our discussion will
revolve mostly around lattice objects, we reserve the standard notation (e.g. $x$, $q(x)$, 
$G(x)$) to represent lattice quantities and/or quantities in lattice units. 
The corresponding physical counterparts will be distinguished by superscript $p$
(e.g. $x^p$, $q^p(x^p)$, $G^p(x^p)$).

\medskip

\noindent
{\bf 2. Reflection Positivity and Locality.}
We will discuss the 2-point function of TChD
\begin{equation}
    G^p(x^p) \,\equiv\, \langle \, q^p(x^p)\, q^p(0) \,\rangle
   \label{eq:5}    
\end{equation}
in Euclidean gauge theory. To show that $G^p(x^p)\le 0$ for $|x^p|>0$ is straightforward. 
Indeed, let us put the origin of the new coordinate system at the midpoint between 
$0$ and $x^p$ and consider a reflection $\theta$ with respect to the axis connecting these 
points, so that $x^p=\theta(0)$. Due to the pseudoscalar nature of $q^p(x^p)$ we then have
$q^p(x^p)=-\Theta q^p(0)$, if $|x^p|>0$. Here $\Theta$ is the (antilinear) reflection 
operator. Consequently
\begin{equation}
    \langle \, q^p(x^p)\, q^p(0) \,\rangle \,=\, 
   -\langle \, \Theta q^p(0) \, q^p(0) \,\rangle \,\le\, 0
   \label{eq:10}    
\end{equation}
where the last inequality follows from reflection positivity of the theory. 
The point $x^p=0$ is singular in this regard and the correlator is obviously positive at the
origin.

   \begin{figure}[t]
   \begin{center}
     \vskip -0.20in
     \centerline{
     \hskip -0.00in
     \includegraphics[height=6.0truecm,angle=0]{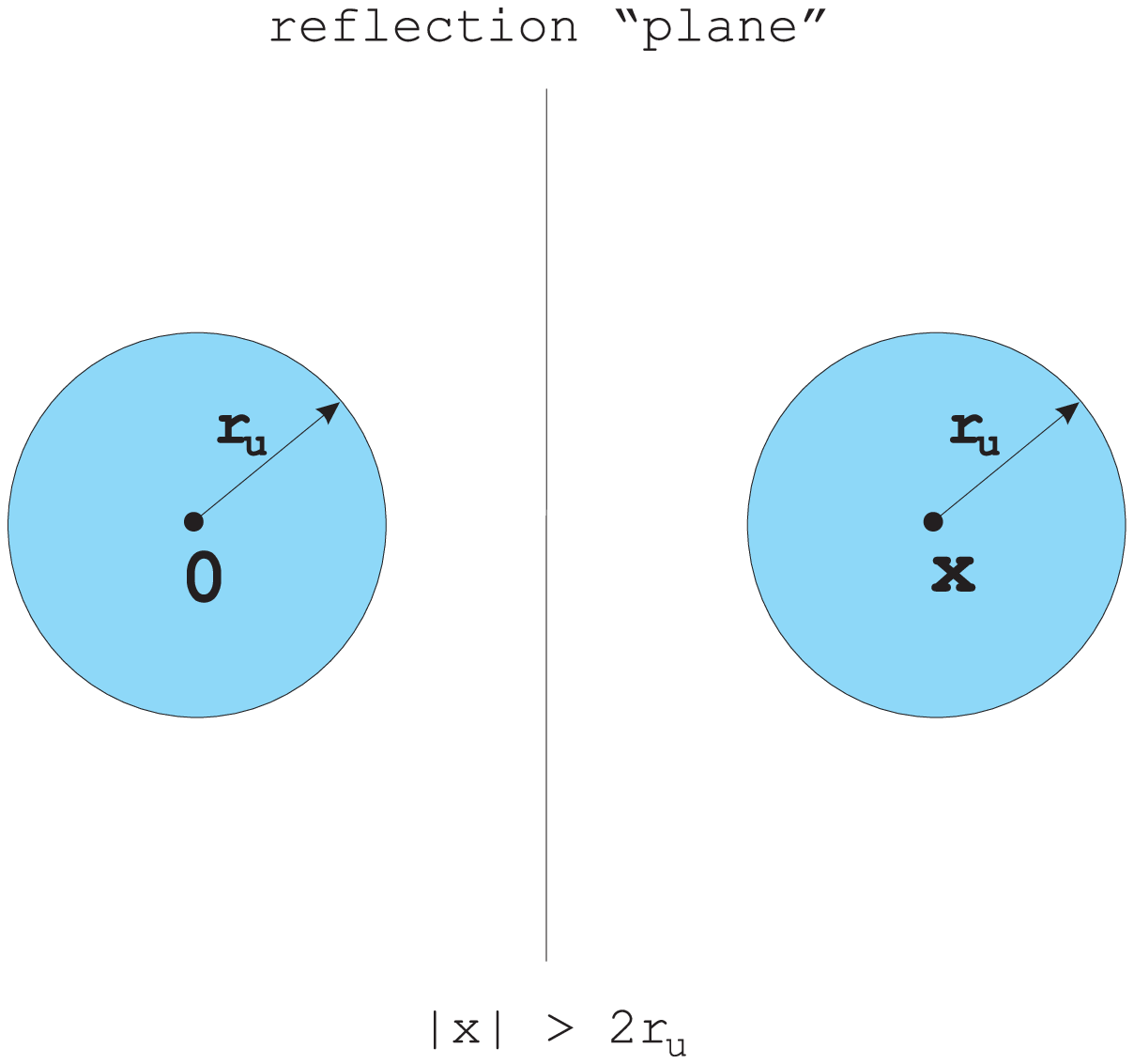}
     \hskip 1.50in
     \includegraphics[height=6.0truecm,angle=0]{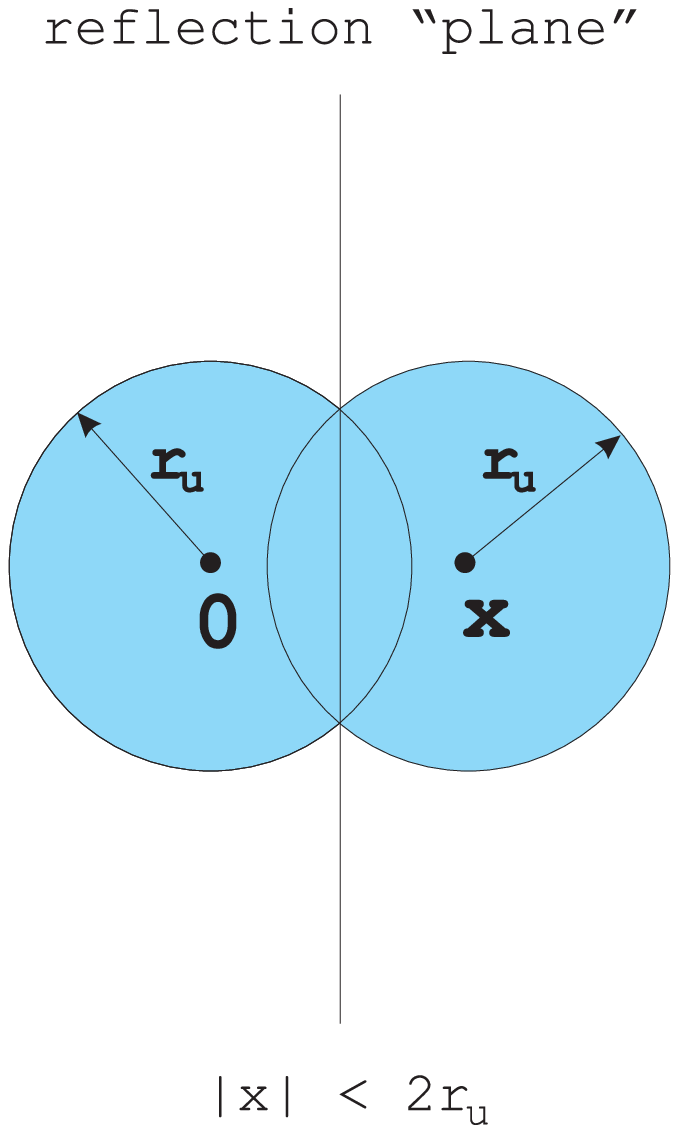}
     }
     \caption{For ultralocal $q(x)$ with range $r_u$ one can directly prove the negativity
              of $G(r,a)$ (in reflection-positive lattice theory) for $|x|> 2r_u$ (left).
              For $|x|< 2r_u$ (right) the operator extends beyond the reflection ``plane'' 
              and the positive core involving non-zero lattice distances may develop in 
              the 2-point function.}
     \vskip -0.4in 
     \label{ref_pos:fig}
   \end{center}
   \end{figure} 

The situation on the lattice is more complicated with complications coming from 
two sources. (1) The lattice theory defined via particular action $S$ may not be strictly 
reflection-positive, i.e. it cannot be proved (via neither site nor link reflection) 
that $\langle \, \Theta F \, F\rangle \ge 0$ for any operator function $F$ depending on field 
variables at {\em arbitrary} positive lattice times. (2) The lattice 
operator $q(x)$ can extend over several lattice spacings, and thus for sufficiently small 
(lattice) $x$ one cannot make the direct argument even if $S$ is strictly reflection positive. 
Indeed, if $q(x)$ is ultralocal with lattice radius $r_u$ then for $|x|<2r_u$ the above 
reflection positivity reasoning breaks down since the operator will extend beyond the 
reflection ``plane''. This is schematically illustrated in Fig.~1. Nevertheless, 
the consequences of reflection positivity will be recovered in the continuum limit since 
the physical size $2 r_u a$ of the (possibly) violating region will go to zero. The situation 
is more involved if operator $q(x)$ is non-ultralocal (such as an operator based on 
Ginsparg-Wilson kernel). In this case the operator {\em always} extends beyond the reflection 
plane and the exact lattice arguments based on reflection positivity cannot be used even 
if $S$ is reflection positive. However, if a non-ultralocal operator is exponentially local 
with the associated finite lattice range $r_{exp}$, then one expects the behavior similar 
to that of an ultralocal operator with comparable range. Specifically, $G(x)$ could contain 
a positive core with radius $r_c \approx r_{exp}$. In fact, one could use the measured size 
of the positive core as a very rough estimate of the lattice range of such an operator 
(assuming that the lattice theory is otherwise reflection positive). If the lattice operator 
is not exponentially local, the negativity of the correlator could be violated at finite 
physical distance in the continuum limit even if the underlying lattice theory is strictly 
reflection positive. If that happens, the corresponding operator should be viewed as 
non-local and discarded.

In later sections we will focus on demonstrating the negativity of $G(x)$ in the continuum
limit of pure-glue QCD defined by the Iwasaki gauge action and using the TChD operator based 
on the overlap Dirac matrix. We will thus be dealing with a lattice theory without strict 
reflection positivity, but with an ultralocal action (with the extent of just two lattice 
spacings) for which there is little doubt that the consequences of reflection positivity 
will hold accurately even before taking the continuum limit. Consequently, the demonstration 
of negativity for $G(x)$ will represent mainly a check on the locality properties
of the overlap-based $q(x)$. In case of very smooth (``admissible'') 
gauge fields the locality of $q(x)$ follows from arguments given in 
Ref.~\cite{ov_loc} for locality of the overlap Dirac operator.\footnote{Note that 
discussion in Ref.~\cite{ov_loc} focuses on locality properties of $D$ in terms of 
fermionic degrees of freedom. What is relevant here is the locality in gauge 
variables.} However, the explicit check of locality for the overlap-based TChD operator 
over realistic lattice QCD ensembles has not been done and remains a very relevant issue. 
Moreover, the question of effective range of the operator (and the size of 
the positive core of the correlator) is of practical interest.

\begin{table}[t]
  \centering
  \begin{tabular}{cccccc}
  \hline\hline\\[-0.4cm]
  \multicolumn{1}{c}{ensemble}  &
  \multicolumn{1}{c}{$a$ [fm]}  &
  \multicolumn{1}{c}{$V$}  &
  \multicolumn{1}{c}{$V^p$ [fm$^4$]} &
  \multicolumn{1}{c}{$\quad$configs$\quad$} \\[2pt]
  \hline\\[-0.4cm]
   $\;\cE_1\;$ & $\quad 0.165 \quad$ & $\quad  8^4 \quad$ & $\quad 3.0 \quad$ 
           & $\quad 50 \quad$\\
   $\;\cE_2\;$ & $\quad 0.110 \quad$ & $\quad 12^4 \quad$ & $\quad 3.0 \quad$ 
           & $\quad 50 \quad$\\
   $\;\cE_3\;$ & $\quad 0.082 \quad$ & $\quad 16^4 \quad$ & $\quad 3.0 \quad$ 
           & $\quad 25 \quad$\\
\hline \hline
\end{tabular}
\caption{Ensembles of Iwasaki gauge configurations for overlap TChD calculation.}
\label{ensemb_tab} 
\end{table}

\medskip
\noindent
{\bf 3. Lattice Data.}
We will work with lattice TChD given by~\cite{Has98A}
\begin{equation}
   q(x) \;=\; \frac{1}{2\rho} \, \mbox{\rm tr} \,\gamma_5 \, D_{x,x} \;\equiv\;  
             -\mbox{\rm tr} \,\gamma_5 \, (1 - \frac{1}{2\rho}D_{x,x})
   \label{eq:15}  
\end{equation}  
where $D$ is the overlap Dirac operator~\cite{Neu98BA} based on the Wilson-Dirac 
kernel with mass $-\rho$. For the numerical results presented here we use the
value $\rho=1.368$ ($\kappa=0.19$). Details of the numerical implementation 
for overlap matrix--vector operation needed to evaluate $q(x)$ can be found in 
Ref.~\cite{Chen03}. The 2-point correlation function was calculated over the 
ensembles of Iwasaki action at three different lattice spacings with details 
specified in Table~\ref{ensemb_tab}. The scale has been determined from string 
tension and the physical size $L^p=1.32$ fm is the same for all ensembles. 
The density $q(x)$ was evaluated at every point of the lattice and, consequently, 
all the correlators computed include contributions from all possible pairs of points
(``all-to-all'' correlators). For ensemble $\cE_1$ a single point source has 
been used to evaluate the density individually for each point. For ensembles 
$\cE_2$ ($\cE_3$) the superposition of 2 (8) maximally separated point sources was 
used to evaluate the density at 2 (8) points simultaneously, thus speeding 
the calculation up accordingly. In case of $\cE_3$ this leads to 
a typical relative error of calculation around $10^{-5}$, and this error is better 
than $2 \times 10^{-3}$ for all but 2\% of the least intense (as measured by 
$|q(x)|$) points. Such deviations have a negligible effect on the correlator. 
The precision is even better for $\cE_1$ and $\cE_2$. 

\begin{figure}
   \begin{center}
     \vskip -0.15in
     \centerline{
     \includegraphics[width=9.5truecm,angle=-90]{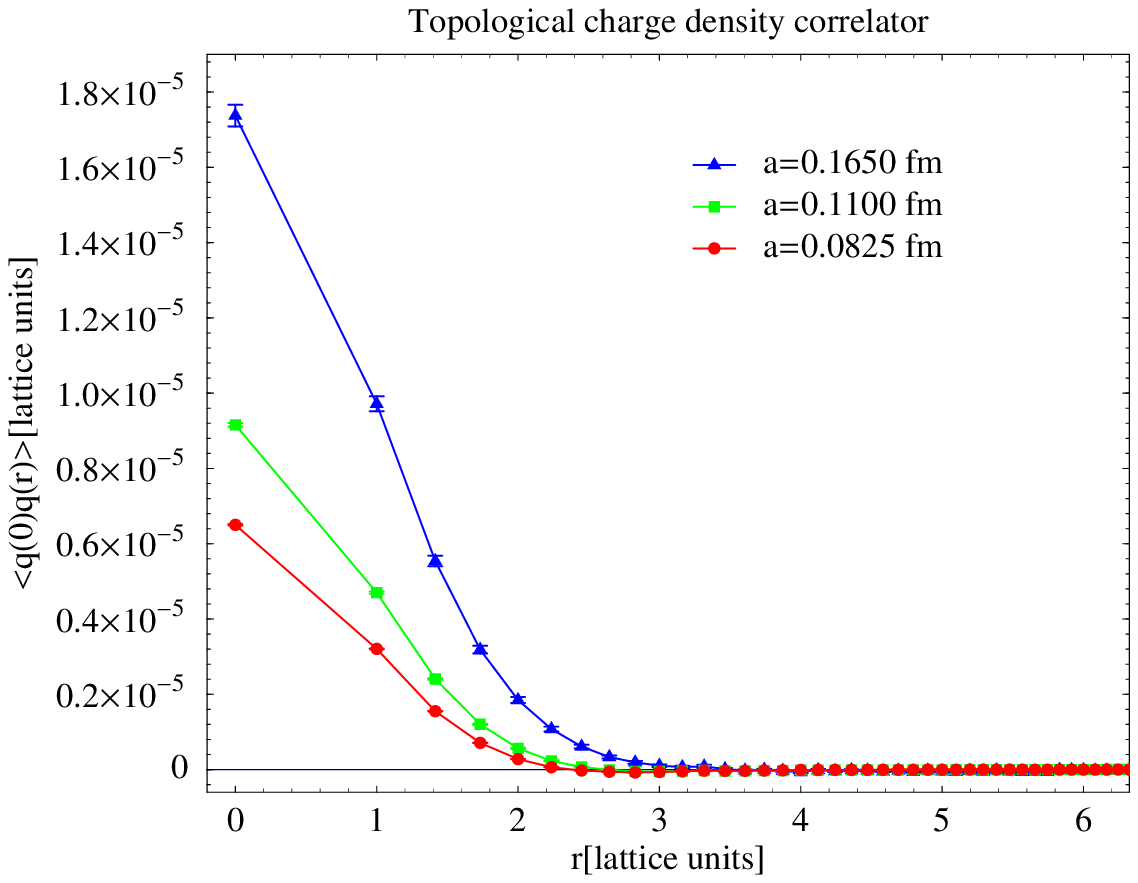}
     }
     \vskip 0.15in
     \centerline{
     \includegraphics[width=9.5truecm,angle=-90]{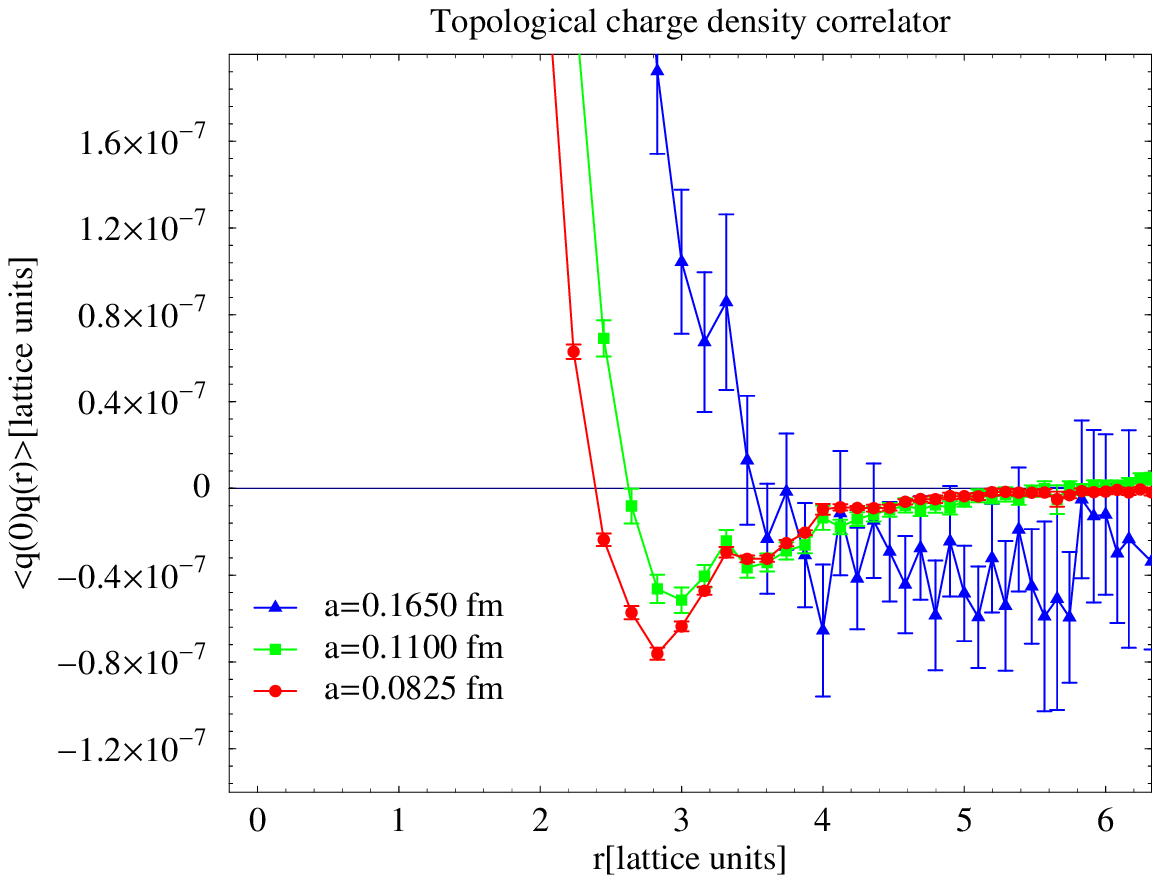}
     }
     \vskip 0.15in
     \caption{Lattice 2-point functions of TChD for ensembles $\cE_1$, $\cE_2$ 
              and $\cE_3$. Details of the behavior for small values of $G(r,a)$ 
              are shown on the bottom plot.}
     \label{lat_cors:fig}
   \end{center}
\end{figure}

Computed lattice correlation functions for ensembles $\cE_1$, $\cE_2$ and $\cE_3$ 
are shown in Fig.~\ref{lat_cors:fig} as a function of $r\equiv |x|$. The bottom part 
of the figure displays the detail of the behavior for small values of $G(r)$. 
One can clearly see that for all three ensembles the correlator exhibits a positive 
core followed by a negative behavior at larger lattice distances. For the analysis 
that will follow we wish to highlight a particular observation indicated by our raw 
lattice data. 

\noindent {\em Observation 1}: The range (width) of the positive core in 
lattice units {\em decreases} as the continuum limit is approached.
\vskip 0.08in
\noindent There are other interesting properties exhibited by the data that we will 
discuss in a forthcoming publication. Here we wish to generalize the above 
observation into the corresponding precise statement which can be verified by further 
simulations. We suggest that the conjecture below is valid at least for the set of 
standard ultralocal gauge actions such as Wilson, Iwasaki, L\"uscher-Weisz, and DBW2 
actions\footnote{Our earlier results with Wilson gauge action~\cite{Hor02Bpr} 
support this conclusion.} and probably much more generally. Also, we expect it 
to be valid for a generic value of $\rho$ (i.e. $0<\rho<2$) in the definition of 
the overlap matrix and the corresponding TChD operator. In arguments that follow
we implicitly assume that a sufficiently large physical volume 
(e.g. larger than 1~fm$^4$) is kept fixed as the lattice spacing is changed toward 
the continuum limit.
\medskip

\noindent {\em Conjecture 1}: There exists a finite lattice spacing $a_0$ 
satisfying the following requirements. (i) For all $a \le a_0$ there is a finite 
lattice distance $r_c(a)$ (``size'' of the positive core) such that 
$G(r,a) \ge 0$ for $r \le r_c(a)$ and $G(r,a) < 0$ for $r > r_c(a)$. 
(ii) The function $r_c(a)$ is {\em non-increasing} with decreasing $a$ for 
$a \le a_0$.\footnote{One would normally say in mathematics that $r_c(a)$ is 
a non-decreasing function in the vicinity of $a=0$. However, in lattice gauge 
theory one usually thinks of changing $a$ from finite values to zero rather than 
vice-versa.}
\medskip

Before we discuss the implications of the above conjecture for the negativity 
of the TChD 2-point function in the continuum limit, we wish to emphasize the (perhaps)
unintuitive nature of {\em Conjecture 1}. Indeed, the standard expectation is that 
the typical gauge fields become ``smoother'' in terms of lattice distances as the 
continuum limit is approached. However, one obviously needs to be careful about the 
interpretation of this expectation since according to {\em Conjecture 1} the typical 
lattice distance over which $q(x)$ changes sign actually shrinks with decreasing 
lattice spacing. This effect has already been noted in Ref.~\cite{Hor03A} where it 
manifested itself via the fact that the size of maximal connected regions built from 
sign-coherent 4-d hypercubes decreases {\em even in lattice units} as the continuum 
limit is approached. This trend is presumably associated with increasingly more 
definite formation of low-dimensional sign-coherent structure in the vacuum closer 
to the continuum limit. Related to this is another trend exhibited by our data
shown in Fig.~\ref{lat_cors:fig}. In particular, there is a definite lattice distance 
$r_d(a)$ (clearly identifiable for $\cE_2$ and $\cE_3$) for which the maximal 
negative value of the correlator in lattice units $G^{min}(a) \equiv \min_r G(r,a)<0$ 
is achieved. The function $r_d(a)$ is non-increasing (similarly to $r_c(a)$) with 
decreasing lattice spacing. Moreover, for the window of lattice spacings studied 
here, the magnitude $-G^{min}(a)$ of maximal anticorrelation {\em increases} with 
decreasing lattice spacing. At the same time 
$G^{max}(a) \equiv \max_r G(r,a) = G(0,a) > 0$ {\em decreases}. In other words, while 
the typical value of $q(0)^2$ decreases closer to the continuum limit, the typical 
magnitude of maximal anticorrelation  $q(0) q(r_d)$ grows in this range of lattice 
spacings.\footnote{We should emphasize that we do not predict the increasing trend for 
$-G^{min}(a)$ to continue arbitrarily close to the continuum limit.}
We again associate this unusual behavior with the presence of a low-dimensional
sign-coherent structure in typical configurations~\cite{Hor03A}, and the fact that
this structure becomes more sharply defined closer to the continuum limit. 
\medskip\smallskip

\noindent
{\bf 4. The Negativity of the Correlator.}
{\em Conjecture 1} has immediate consequences relevant for the negativity of the 
overlap-based TChD correlator in the continuum limit.
\vskip 0.1in
\noindent {\em Corollary 1}: The size of the positive core in physical units
$r_c^p(a) \equiv a r_c(a)$ vanishes in the continuum limit.
\vskip 0.1 in
\noindent Indeed, according to {\em Conjecture 1} we have 
\begin{displaymath}
  r_c^p(a) \,=\, a r_c(a) \,\le\, a r_c(a_0) \; \rightarrow \; 0 
 \quad \mbox{\rm for} \quad a \; \rightarrow \; 0     
\end{displaymath}
We thus conclude that the TChD 2-point function obtained as a continuum limit of 
the lattice correlator using an overlap-based TChD operator is negative at arbitrary 
non-zero distances as required by reflection positivity arguments in the continuum.

\begin{figure}
   \begin{center}
     \vskip -0.10in
     \centerline{
     \hskip -0.02in
     \includegraphics[width=6.2truecm,angle=-90]{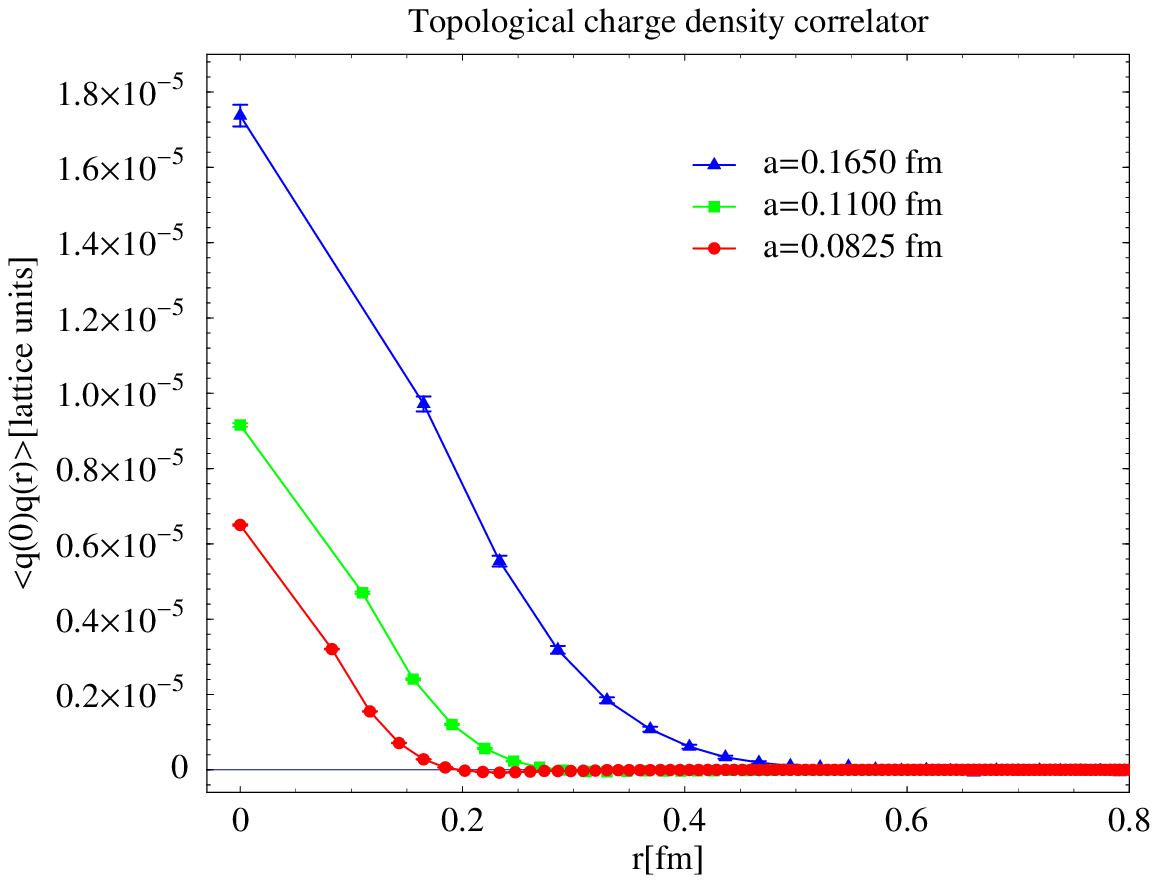}
     \hskip -0.22in
     \includegraphics[width=6.2truecm,angle=-90]{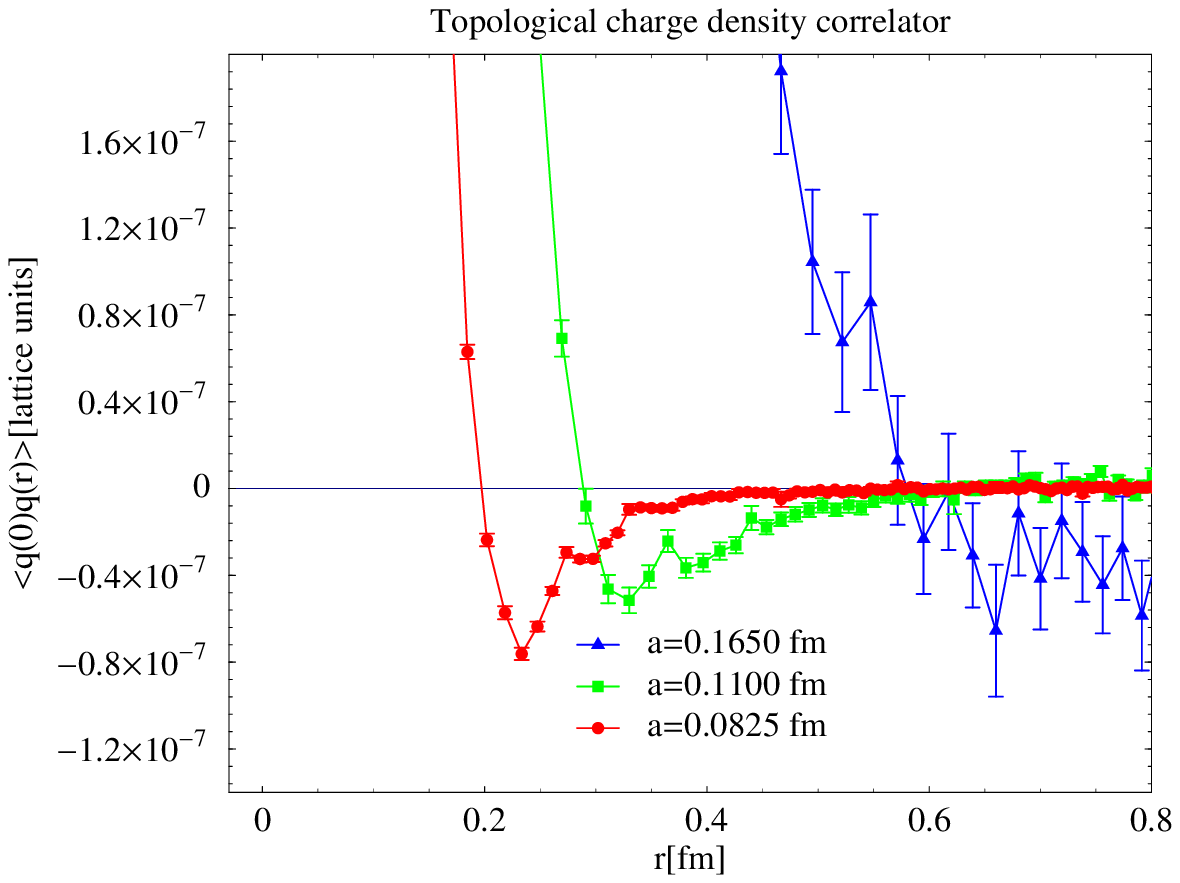}
     }
     \vskip 0.10in
     \centerline{
     \hskip -0.06in
     \includegraphics[width=6.2truecm,angle=-90]{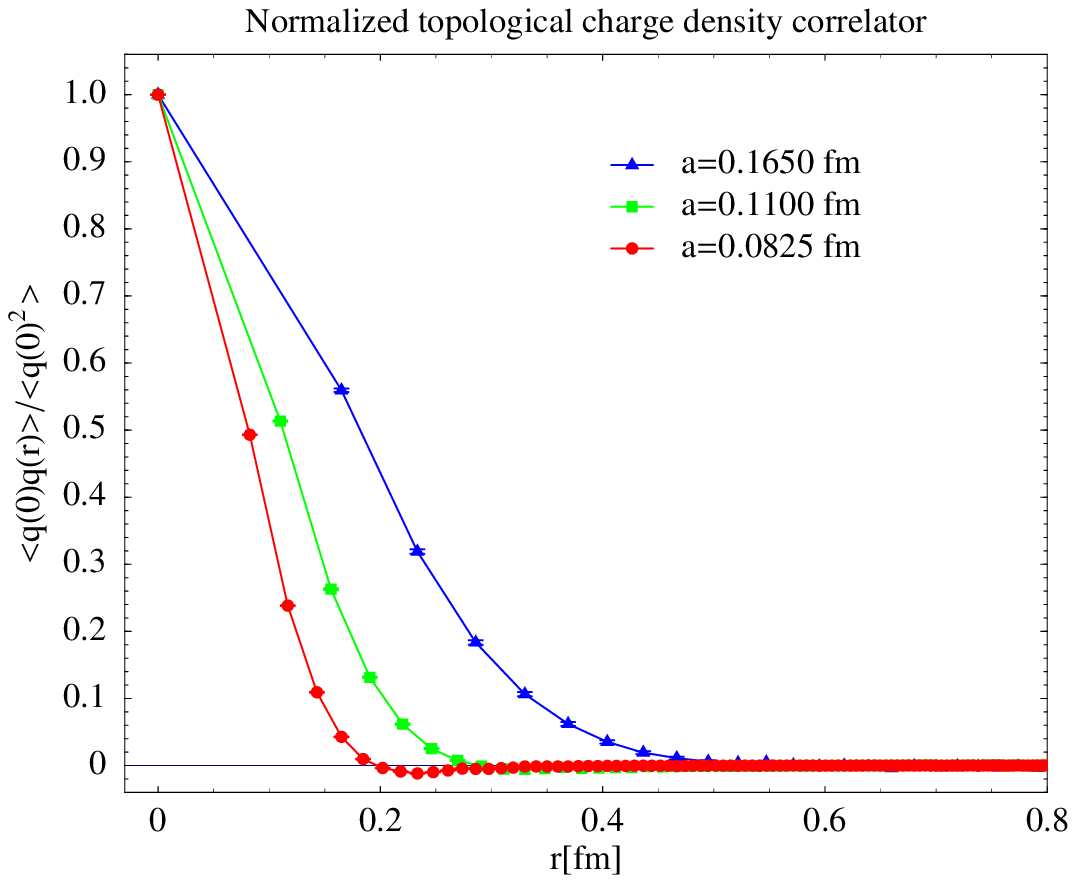}
     \hskip -0.25in
     \includegraphics[width=6.2truecm,angle=-90]{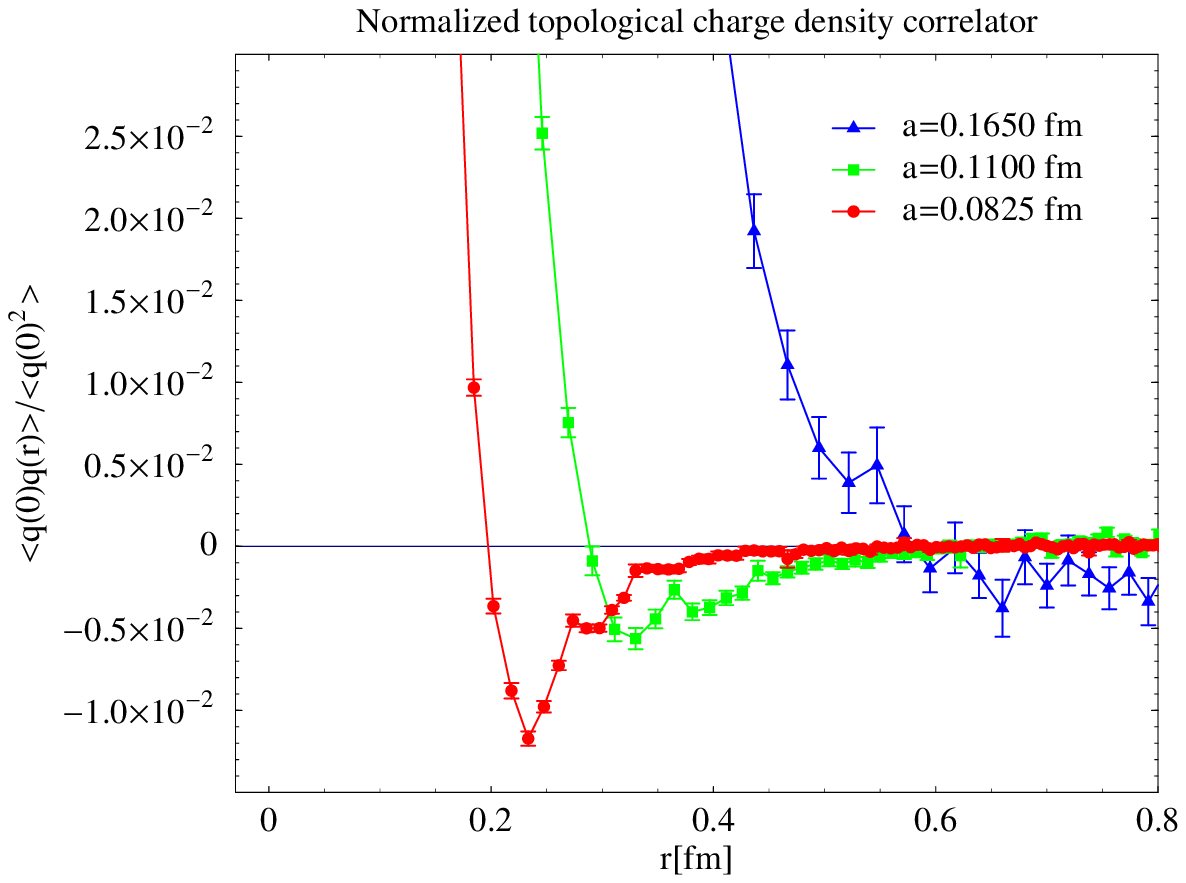}
     }
     \vskip 0.10in
     \centerline{
     \hskip -0.06in
     \includegraphics[width=6.2truecm,angle=-90]{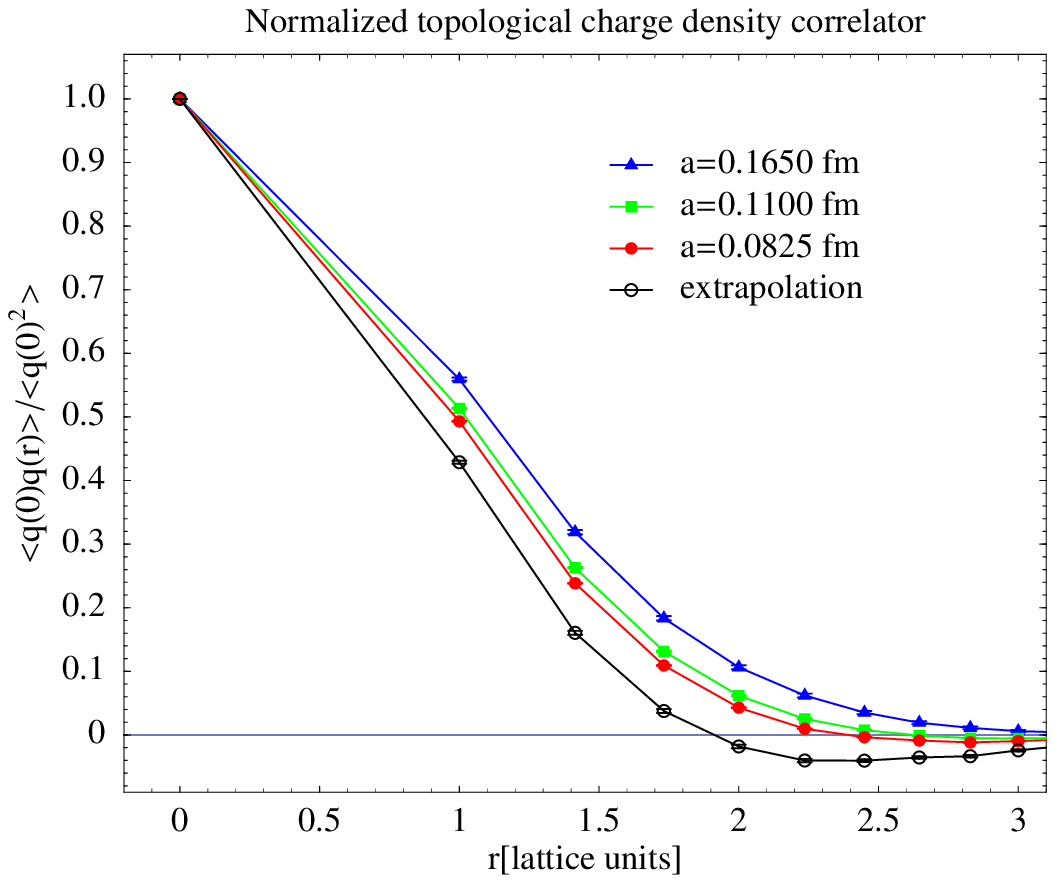}
     \hskip -0.25in
     \includegraphics[width=6.2truecm,angle=-90]{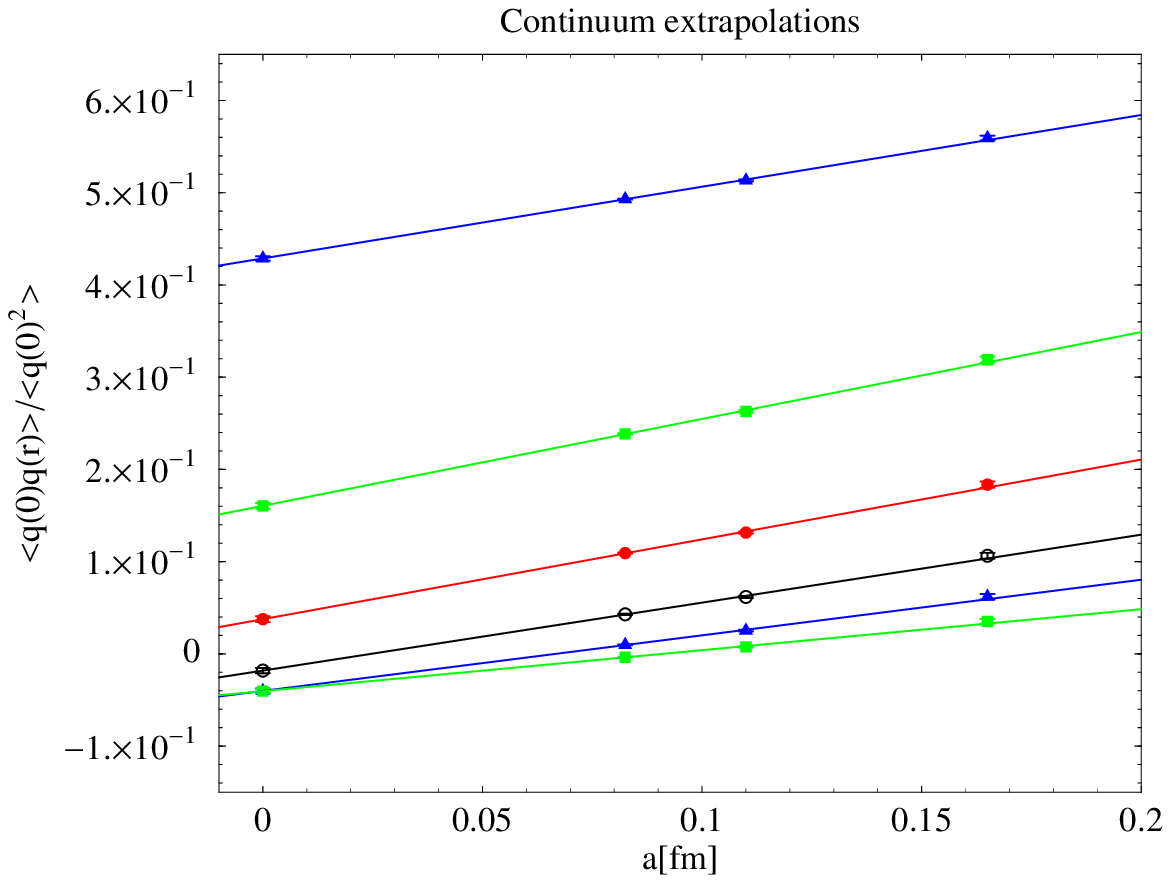}
     }
     \caption{(Top) Lattice 2-point functions of TChD from Fig.~\ref{lat_cors:fig}
              plotted against physical distance.
              (Middle) 2-point function normalized at the origin versus physical distance.
               The detail is on the right. 
              (Bottom) Short (lattice) distance behavior of $G^N(r,a)$ together with 
              the continuum extrapolation is shown on the left. The individual 
              extrapolations for $r^2=1,\ldots,6$ (in this order from top down) are shown 
              on the right.}
     \label{core:fig}
   \end{center}
   \end{figure}

To see pictorially how the size of the positive core in physical units decreases as 
the lattice spacing is lowered, we show the computed lattice correlation functions 
against physical distance in Fig.~\ref{core:fig} (top) with detail of the negative 
behavior shown on the right. The shrinking of the positive core toward zero 
physical range can be most clearly seen by separating the lattice-spacing dependence
of the shape of the correlator from that of its magnitude. In other words, we write
\begin{equation}
   G(r,a) \,\equiv\, G(0,a) \, G^N(r,a)
   \label{eq:20}
\end{equation}
where $G^N(r,a)$ (the ``shape'') is normalized to unity at the origin. The behavior
of $G^N(r^p,a)$ is shown in Fig.~\ref{core:fig} (middle) with detail on the right. 
We note that the errorbars on these correlators were determined by applying 
the jacknife procedure directly to $G^N(r^p,a)$ and thus are zero at the origin.  

We now wish to use our lattice data to extract more detailed information on how 
the size of the positive core behaves close to the continuum limit. {\em Conjecture 1} 
straightforwardly implies the following statement.
\vskip 0.1in
\noindent {\em Corollary 2}: The function $r_c(a)$ has a well-defined non-diverging 
continuum limit, i.e.
\begin{displaymath}
        0 \,\le\, \lim_{a\to 0}\, r_c(a) \,\equiv\, r_c(0) \,<\, \infty 
\end{displaymath}
\vskip -0.06in
\noindent Moreover, there exists a non-zero lattice spacing $a_c<a_0$ such that 
$r_c(a)=r_c(0)$ for $a \le a_c$. 

\vskip 0.1in
\noindent
Indeed, since $r_c(a)$ is bounded from below and non-increasing as $a \rightarrow 0$, 
the limit is guaranteed to exist. Moreover, $r_c(a)$ is a discrete-valued function 
with possible values such that $r_c^2 \in \{ 0,1,2,\ldots \}$ (non-negative integers). 
This means that there is only a finite number of possible values smaller than $r_c(a_0)$ 
thus implying the second part of the statement. Our goal is to determine $r_c(0)$. 
To do that, it is actually practical to attempt a more general calculation. 
In particular, we will determine the {\em shape} of the correlator (i.e. $G^N(r,a)$) 
at short lattice distances and arbitrarily close to the continuum limit. More precisely, 
we will assume that the point-wise continuum limit 
$\lim_{a \to 0} G^N(r,a)\equiv G^N(r)$ exists, and that $G^N(r,a)$ can be 
power-expanded around it for arbitrary fixed lattice $r$, i.e.
\begin{equation}
   G^N(r,a) \,=\, G^N(r) \,+\, \sum_{k=1}^{\infty} a^k G^N_k(r)
   \label{eq:25}
\end{equation}
Here $G^N_k(r)$ are finite functions (with values depending on the choice
of units for $a$). The convergence is expected to be non-uniform across the
domain of $r$. The correlators $G^N(r,a)$ for small lattice distances are shown
in Fig.~\ref{core:fig} (bottom, left). The lattice spacing dependence of 
$G^N(r,a)$ for $r^2=1,\ldots,6$ is shown on the right, indicating that the linear 
terms in the expansion (\ref{eq:25}) are dominant for this range of lattice 
spacings/lattice distances. We thus perform a linear extrapolation to the 
continuum limit to estimate $G^N(r)$. The result of the extrapolation is shown 
together with normalized correlators. One can simply read off the plot that 
$r_c(0)=\sqrt{3}$. We need to point out here that while the statistical significance 
of this result is very good (the errorbars are barely visible), there will be a small 
systematic effect present due to the fact that in the continuum extrapolation we 
neglect higher orders in lattice spacing. Indeed, a very small positive curvature 
can be seen under close inspection of Fig.~\ref{core:fig} (bottom, right). This
will lead to a small shift of the extrapolated values around $r_c(0)$ in the 
upward direction. A simple estimate from the curvature gives a correction such 
that $G^N(r=2)\approx 0$. We thus conclude that 
$r_c(0)=\sqrt{3}$ or $2$ and that sufficiently close to the continuum limit
the size of the positive core is $r^p_c \approx 2 a$.  

It should be emphasized that contrary to the content of {\em Conjecture 1}, 
we do not expect the specific value of $r_c(0)$ to be strictly universal with 
respect to the set of ``standard'' gauge actions or the family of TChD operators 
labeled by mass parameter $\rho$. For example, changing $\rho$ can probably 
have some effect on $r_c(0)$ since the precise lattice localization range of 
the corresponding TChD operator can change~\cite{ov_loc}. 
\medskip

\noindent
{\bf 5. Divergent Contact Part.}
Finally, we wish to illustrate how the expected non-integrable nature of 
the contact part in TChD 2-point function manifests itself in the regularized
correlator as the continuum limit is approached. Let us first recall the
standard expectation that $q^p(x^p)$ is a dimension 4 operator (i.e. has no 
anomalous dimension) and thus $G^p(x^p)$ should behave near the origin as 
$\sim -|x^p|^{-8}$ up to possible logarithms. Such a singularity is clearly non-integrable and 
thus, in order to obtain a finite positive space-time integral (susceptibility), 
the appropriate non-integrable counterterms must be present with support
at the origin~\cite{SeSt}. In particular, the contact part is expected
to have the form~\cite{SeSt,Giu02A}
\begin{equation}
  c_1 \, \delta (x^p) + c_2 \, \Delta \delta (x^p) + c_3\, \Delta^2 \delta (x^p)
  \label{eq:30}
\end{equation}
with $\Delta$ denoting the Laplacian and $c_1$, $c_2$ and $c_3$ being free
parameters.\footnote{The fact that these parameters do not seem to be fixed 
by general considerations while topological susceptibility depends on $c_1$
is one of the intriguing aspects of this subject.} 

   \begin{figure}[t]
   \begin{center}
     \vskip -0.20in
     \centerline{
     \hskip -0.00in
     \includegraphics[width=6.2truecm,angle=-90]{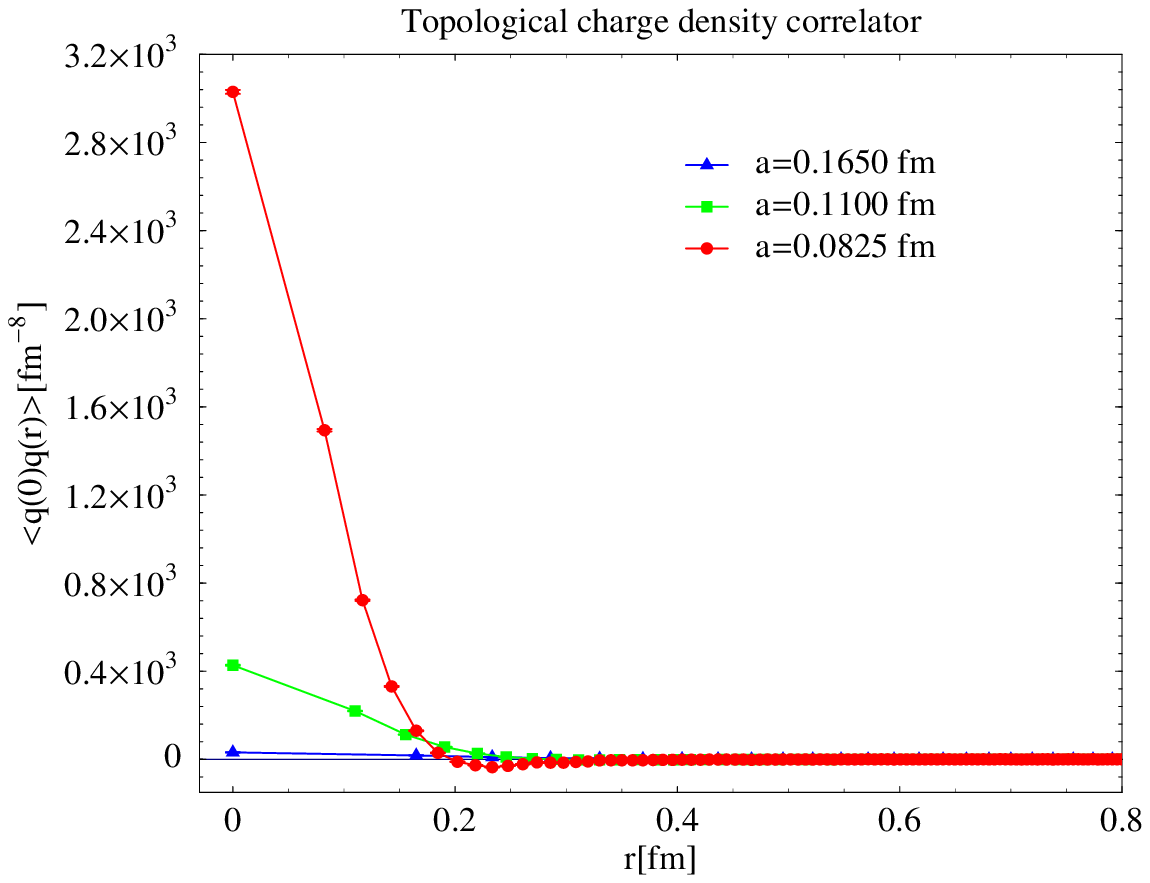}
     \hskip -0.35in
     \includegraphics[width=6.2truecm,angle=-90]{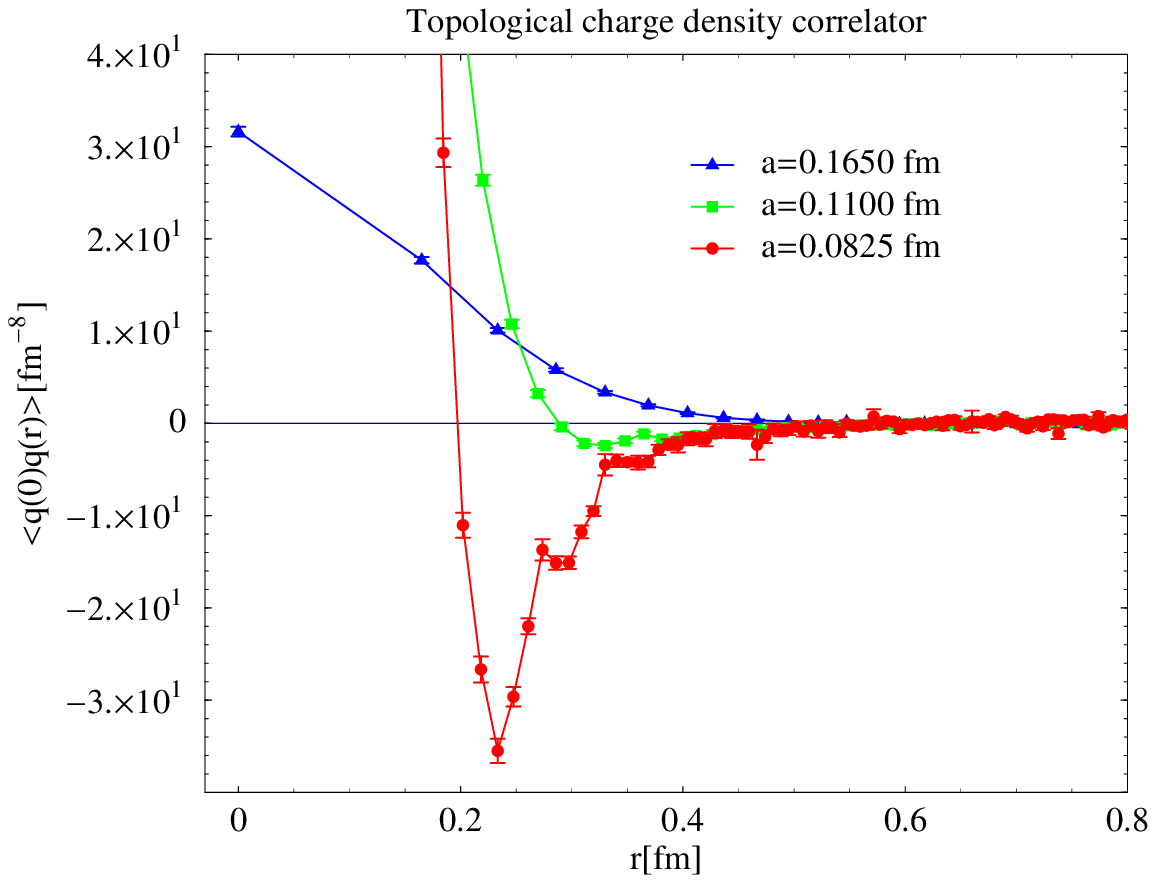}
     }
     \caption{TChD 2-point functions in physical units. The detail of the negative 
              behavior is shown on the right. Data indicate the divergent 
              nature of both the positive core and the negative dip.}
     \vskip -0.4in 
     \label{phy_cors:fig}
   \end{center}
   \end{figure} 

\begin{figure}[t]
   \begin{center}
     \vskip -0.20in
     \centerline{
     \hskip -0.00in
     \includegraphics[width=9.0truecm,angle=-90]{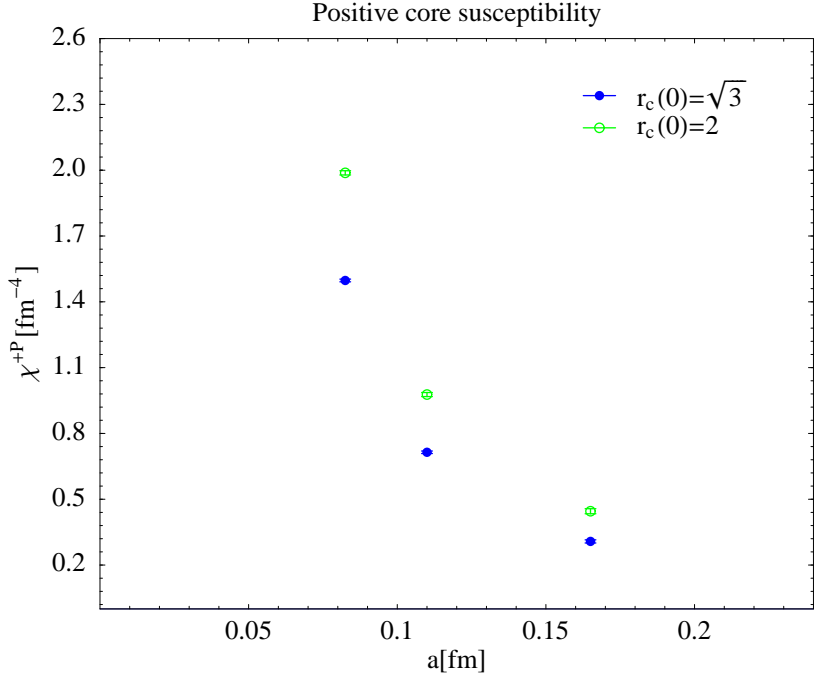}
     }
     \caption{The contribution of the positive core to the topological susceptibility
              using $r_c(0)=\sqrt{3}$ and $r_c(0)=2$. Data indicates the diverging 
              behavior in the continuum limit.}
     \vskip -0.4in 
     \label{core_susc:fig}
   \end{center}
   \end{figure}

In the ideal situation, all of the above statements should be verified using 
the measured lattice correlator $G(r,a)$ as the sole input. Here we restrict
ourselves to illustrating that the {\em non-integrable} positive core indeed
emerges in the lattice definition of the theory with overlap-based TChD 
operator. We will thus simply assume that the finite continuum limit 
\begin{equation}
    G^p(r^p) = \lim_{a \to 0} G^p(r^p,a) = 
               \lim_{a \to 0} \frac{G(r^p/a,a)}{a^8}
   \label{eq:35}
\end{equation}
exists for arbitrary $r^p > 0$. To monitor what happens at $r^p=0$ (i.e. 
in the lattice positive core which shrinks to $r^p=0$ in physical units)
as the continuum limit is approached, we plot $G^p(r^p,a)$ for our ensembles 
in Fig.~\ref{phy_cors:fig}. One can see that the emergence of the divergent 
positive core is very dramatic as the continuum limit is approached. The precise 
nature of this divergence will be discussed quantitatively in a forthcoming 
publication. To see that the singularity is indeed non-integrable we compute 
the contribution of the positive core to susceptibility, namely
\begin{equation}
    \chi^{+p}(a) = a^{-4} \sum_{r \le r_c(0)} G(r,a) N(r) 
   \label{eq:40}
\end{equation}
where the factor $N(r)$ represents the multiplicity of points $x$ on the hypercubic 
lattice such that $|x|=r$. Using both $r_c(0)=\sqrt{3}$ and $r_c(0)=2$ (see discussion 
in the previous section) we plot $\chi^{+p}(a)$ for our ensembles in 
Fig.~\ref{core_susc:fig}. The data clearly exhibits the convex increasing behavior 
of $\chi^{+p}(a)$ when approaching the continuum limit in agreement with the expected 
non-integrable nature of the singularity in $G^p(r^p)$. In order to obtain finite 
positive susceptibility,\footnote{High statistics numerical evidence confirming that 
the susceptibility defined via overlap-based TChD operator is indeed finite can be 
found in Refs.~\cite{Giu03,DelDeb04}. Our data is consistent with this finding.} 
the diverging contribution of the positive core to the integral of $G^p(x^p)$ has 
to be canceled by the opposite divergence in the negative part. The emergence
of divergent negative dip as the continuum limit is approached is clearly visible 
in Fig.~\ref{phy_cors:fig} (right). 
\medskip

{\bf 6. Conclusions.}
We have performed the first calculation of the lattice 2-point function of TChD 
in pure-glue QCD using the topological field operator based on the overlap Dirac 
matrix. In this initial study, we have focused on basic properties of the 
correlator expected on general grounds.\footnote{Very recently a manuscript dealing 
with related issues has been released~\cite{Agua05}.} Thus, our main motivation 
was to check whether an overlap-based TChD operator offers a valid definition of 
the topological field in the continuum limit. Our main conclusions are the following.
\medskip

\noindent {\em (i)} The correlator exhibits a positive core over finite lattice 
distance $r_c(a)$ and is negative for larger lattice distances. The function
$r_c(a)$ is {\em non-increasing} as the continuum limit is approached, implying 
that the size of the positive core in physical units is zero in the continuum
limit. Consequently, the overlap-based TChD operator complies with the requirement
imposed by reflection positivity. This result also indicates, in an indirect manner, 
that the overlap-based TChD operator is {\em local\/} for ensembles used in 
realistic lattice simulations.
\medskip

\noindent {\em (ii)} Data presented in this article indicate that the value of 
the correlator at the origin diverges in physical units as does the value at 
the maximal negative ``dip'' of the correlator. Moreover, the contribution of 
the positive core to susceptibility also exhibits divergent behavior thus
indicating the presence of a positive {\em non-integrable} contact part. 
This is in agreement with conclusions obtained by formal 
considerations~\cite{SeSt,Giu02A}. 
\medskip

Let us finally remark that we find it quite intriguing to explicitly see how 
the two infinities (the negative one coming from strong power-law behavior  
and the positive one due to the contact part) manifest themselves 
in the regularized version of the theory. Since the contact part has strict 
support at the origin in the continuum, it is natural to expect that both 
infinities should coexist in the lattice correlator, and emerge simultaneously 
(via lattice point splitting) as the continuum limit is approached. 
The diverging positive core and the diverging negative dip of 
Fig.~\ref{phy_cors:fig} illustrate how this happens in the overlap-based TChD 
correlator.

\bigskip
\noindent
{\bf Acknowledgments:} 
This work was supported in part by U.S. Department of Energy under grants 
DE-FG05-84ER40154 and DE-FG02-95ER40907.

\end{document}
\bye